\documentclass{article}
\usepackage{cite}

\begin{document}

\date{}
\title{Algebraic treatment of non-Hermitian quadratic Hamiltonians}
\author{Francisco M. Fern\'{a}ndez \thanks{%
E-mail: fernande@quimica.unlp.edu.ar} \\
INIFTA, Divisi\'on Qu\'imica Te\'orica\\
Blvd. 113 S/N, Sucursal 4, Casilla de Correo 16, 1900 La Plata, Argentina}
\maketitle

\begin{abstract}
We generalize a recently proposed algebraic method in order to treat
non-Hermitian Hamiltonians. The approach is applied to several quadratic
Hamiltonians studied earlier by other authors. Instead of solving the
Schr\"{o}dinger equation we simply obtain the eigenvalues of a suitable
matrix representation of the operator. We take into account the existence of
unitary and antiunitary symmetries in the quantum-mechanical problem.
\end{abstract}

\section{Introduction}

\label{sec:intro}

Hamiltonian operators that are quadratic functions of the coordinates and
momenta proved to be useful for the study of interesting physical phenomena%
\cite{SLZEK11,RSEGK12,BGOPY13,POLMGLFNBY14}. The eigenvalues of such
Hamiltonians may be real or complex. The occurrence of real or complex
spectrum depends on the values of the model parameters that determine the
experimental setting. The transition from one regime to the other is
commonly interpreted as the breaking of PT symmetry. In some cases those PT
symmetric Hamiltonians are also Hermitian\cite{F15a,F16}.

In addition to those Hamiltonians directly related to experiment there are
other quadratic oscillators that have been used to illustrate physical
concepts in a more theoretical setting. They may be Hermitian\cite
{E80,BG15,RCR14} or non-Hermitian\cite{CIN10,LM12,FG14b,BKB15,MX16}.

The eigenvalue equation for a quadratic Hamiltonian can be solved exactly in
several different ways\cite{RK05,BGOPY13,BG15}. The algebraic method\cite
{F15a,F16}, based on well known properties of Lie algebras\cite{G74,FC96},
is particularly simple and straightforward. It focusses on the natural
frequencies of the quantum-mechanical problem and reveals the transition
from real to complex spectrum without solving the eigenvalue equation
explicitly or writing the Hamiltonian in diagonal form. The whole problem
reduces to diagonalizing a $2N\times 2N$ matrix, where $N$ is the number of
coordinates.

Those earlier applications of the algebraic method focused on Hermitian
Hamiltonians\cite{F15a,F16} but the approach can also be applied to
non-Hermitian quadratic ones. The purpose of this paper is to generalize
those results and take into account possible unitary and antiunitary
symmetries of the quadratic Hamiltonians.

In section~\ref{sec:unitary_anti_sym} we briefly address the concepts of
unitary and antiunitary symmetries. In section~\ref{sec:algebraic_method} we
outline the main ideas of the algebraic method and derive the regular or
adjoint matrix representation for non-Hermitian quadratic Hamiltonians. The
approach is similar to that in the previous papers\cite{F15a,F16} but the
results are slightly more general and convenient for present aims. In this
section we also consider Hermitian quadratic Hamiltonians and illustrate the
main results by means of a simple one-dimensional example. In section~\ref
{sec:QH2D} we consider three two-dimensional quadratic Hamiltonians already
studied earlier by other authors. They prove useful for illustrating the
transition from real to complex spectrum, when the model is either Hermitian
or non-Hermitian, which depends on a suitable choice of the model
parameters. Finally, in section~\ref{sec:conclusions} we summarize the main
results and draw conclusions.

\section{Unitary and antiunitary symmetry}

\label{sec:unitary_anti_sym}

In this section we outline some well known concepts and definitions that
appear in most textbooks on quantum mechanics\cite{CDL77} with the purpose
of facilitating the discussion in subsequent sections. Given a linear
operator $H$ its adjoint $H^{\dagger }$ satisfies
\begin{equation}
\left\langle f\right| H^{\dagger }\left| f\right\rangle =\left\langle
f\right| H\left| f\right\rangle ^{*},  \label{eq:Adjoint}
\end{equation}
for any vector $\left| f\right\rangle $ in the complex Hilbert space where $%
H $ is defined. If $H=H^{\dagger }$ we say that the operator is Hermitian%
\cite{CDL77}.

If $\left| \psi \right\rangle $ is an eigenvector of the Hermitian operator $%
H$ with eigenvalue $E$
\begin{equation}
H\left| \psi \right\rangle =E\left| \psi \right\rangle ,\;
\label{eq:H_psi=E_psi}
\end{equation}
then $\left\langle \psi \right| H\left| \psi \right\rangle =\left\langle
\psi \right| H\left| \psi \right\rangle ^{*}$ leads to $\left(
E-E^{*}\right) \left\langle \psi \right| \left. \psi \right\rangle =0$.
Therefore, if $\left| \psi \right\rangle $ belongs to the Hilbert space
where $H$ is defined ($0<\left\langle \psi \right| \left. \psi \right\rangle
<\infty $) then $E$ is real.

A unitary operator $S$ satisfies
\begin{equation}
\left\langle Sf\right. \left| Sg\right\rangle =\left\langle f\right. \left|
g\right\rangle ,  \label{eq:S_unitary}
\end{equation}
for any pair of vectors $f$ and $g$ in the Hilbert space where $S$ is
defined. If
\begin{equation}
SH=HS,  \label{eq:SH=HS}
\end{equation}
we say that the linear operator $H$ exhibits a unitary symmetry (we assume
that both $H$ and $S$ are defined on the same Hilbert space). We can also
write this expression as $SHS^{\dagger }=H$ because $S^{\dagger }=S^{-1}$.
It follows from equations (\ref{eq:H_psi=E_psi}) and (\ref{eq:SH=HS}) that
\begin{equation}
HS\left| \psi \right\rangle =SH\left| \psi \right\rangle =ES\left| \psi
\right\rangle .  \label{eq:HSPsi}
\end{equation}

An antiunitary operator $A$ satisfies\cite{W60}
\begin{equation}
\left\langle Af\right. \left| Ag\right\rangle =\left\langle f\right. \left|
g\right\rangle ^{*},  \label{eq:A_antiunitary}
\end{equation}
for any pair of vectors $f$ and $g$ in the Hilbert space where $A$ is
defined. It follows from this expression that
\begin{equation}
A\left( af+bg\right) =a^{*}Af+b^{*}Ag,  \label{eq:A_antilinear}
\end{equation}
for any pair of complex numbers $a$ and $b$ and we say that $A$ is antilinear%
\cite{W60}.

If
\begin{equation}
HA=AH,  \label{eq:HA=AH}
\end{equation}
we say that the linear operator $H$ exhibits an antiunitary symmetry (we
assume that $H$ and $A$ are defined on the same Hilbert space). Since $A$ is
invertible this last expression can be rewritten as $AHA^{-1}=H$. An
important consequence of this equation is that
\begin{equation}
HA\left| \psi \right\rangle =AH\left| \psi \right\rangle =AE\left| \psi
\right\rangle =E^{*}A\left| \psi \right\rangle .  \label{eq:HAPsi}
\end{equation}
Therefore, if $A\left| \psi \right\rangle =a\left| \psi \right\rangle $
(that is to say, the antiunitary symmetry is exact) the eigenvalue $E$ is
real (even if $H$ is non-Hermitian).

\section{The algebraic method}

\label{sec:algebraic_method}

In two earlier papers we applied the algebraic method to a class of
Hermitian Hamiltonians that includes those that are quadratic functions of
the coordinates and their conjugate momenta\cite{F15a,F16}. In what follows
we apply the approach to non-Hermitian Hamiltonians that are also of
remarkable physical interest. Although most of the expressions are similar
to those derived in the earlier articles we repeat the treatment here in
order to make this paper sufficiently self contained. The main difference is
that we do not assume that $H$ is Hermitian and the main results will be
more general. In addition to it, present algebraic approach will take into
account the possibility that $H$ exhibits unitary or antiunitary symmetries.

The algebraic method enables us to solve the eigenvalue equation for a
linear operator $H$ in the particular case that there exists a set of
Hermitian operators $S_{N}=\{O_{1},O_{2},\ldots ,O_{N}\}$ that satisfy the
commutation relations
\begin{equation}
\lbrack H,O_{i}]=\sum_{j=1}^{N}H_{ji}O_{j}.  \label{eq:[H,Oi]}
\end{equation}
Without loss of generality we assume that the operators in $S_{N}$ are
linearly independent so that the only solution to
\begin{equation}
\sum_{j=1}^{N}d_{j}O_{j}=0,  \label{eq:lin_indep_cond}
\end{equation}
is $d_{i}=0$ for all $i=1,2,\ldots ,N$.

Because of equation (\ref{eq:[H,Oi]}) it is possible to find an operator of
the form
\begin{equation}
Z=\sum_{i=1}^{N}c_{i}O_{i},  \label{eq:Z}
\end{equation}
such that
\begin{equation}
\lbrack H,Z]=\lambda Z.  \label{eq:[H,Z]}
\end{equation}
The operator $Z$ is important for our purposes because $Z\left| \psi
\right\rangle $ is eigenvector of $H$ with eigenvalue $E+\lambda $:
\begin{equation}
HZ\left| \psi \right\rangle =ZH\left| \psi \right\rangle +\lambda Z\left|
\psi \right\rangle =(E+\lambda )Z\left| \psi \right\rangle ,
\label{eq:HZ|Psi>}
\end{equation}
provided that $Z\left| \psi \right\rangle $ is nonzero.

It follows from equations (\ref{eq:[H,Oi]}), (\ref{eq:Z}) and (\ref{eq:[H,Z]}%
), and from the fact that the set $S_{N}$ is linearly independent, that the
coefficients $c_{i}$ are solutions to
\begin{equation}
(\mathbf{H}-\lambda \mathbf{I})\mathbf{C}=0,  \label{eq:(H-lambda_I)C=0}
\end{equation}
where $\mathbf{H}$ is an $N\times N$ matrix with elements $H_{ij}$, $\mathbf{%
I}$ is the $N\times N$ identity matrix, and $\mathbf{C}$ is an $N\times 1$
column matrix with elements $c_{i}$. $\mathbf{H}$ is called the adjoint or
regular matrix representation of $H$ in the operator basis set $S_{N}$\cite
{G74,FC96}. Equation (\ref{eq:(H-lambda_I)C=0}) admits nontrivial solutions
for those values of $\lambda $ that are roots of the characteristic
polynomial
\begin{equation}
P(\lambda )=\det (\mathbf{H}-\lambda \mathbf{I})=0.  \label{eq:charpoly}
\end{equation}
In some cases the matrix $\mathbf{H}$ may not be diagonalizable because it
is not normal
\begin{equation}
\mathbf{HH}^{\dagger }\neq \mathbf{H}^{\dagger }\mathbf{H}.
\label{eq:H_nonnormal}
\end{equation}

If there exists a linear operator $W$ that commutes with all the basis
operators $O_{i}$ ($[W,O_{i}]=0$, $i=1,2,\ldots ,N$) then $H$ and $H+W$
share the same adjoint matrix $\mathbf{H}$. Such an occurrence is not a
serious difficulty in the cases studied here because any operator $W$ with
such a property is proportional to the identity operator and its effect on
the spectrum is trivial.

Of particular interest for the present paper is a basis set of operators
that satisfy
\begin{equation}
\lbrack O_{i},O_{j}]=U_{ij}\hat{1},  \label{eq:[Oi,Oj]}
\end{equation}
where $U_{ij}$ is a complex number and $\hat{1}$ is the identity operator
that we will omit from now on. It follows from $[O_{j},O_{i}]=-[O_{i},O_{j}]$
and $[O_{i},O_{j}]^{\dagger }=-[O_{i},O_{j}]$ that
\begin{equation}
U_{ij}=-U_{ij}^{*}=-U_{ji},  \label{eq:Uij=-Uji}
\end{equation}
and
\begin{equation}
\mathbf{U}^{\dagger }=\mathbf{U,}  \label{eq:U^dagger=U}
\end{equation}
where $\mathbf{U}$ is the $N\times N$ matrix with elements $U_{ij}$. The
well known Jacobi identity
\begin{equation}
\lbrack O_{k},[H,O_{i}]]+[O_{i},[O_{k},H]]+[H,[O_{i},O_{k}]]=0,
\end{equation}
leads to
\begin{equation}
\lbrack O_{k},[H,O_{i}]]=[O_{i},[H,O_{k}]].  \label{eq:Jac_ident}
\end{equation}
Therefore, equations (\ref{eq:[H,Oi]}), (\ref{eq:[Oi,Oj]}), (\ref
{eq:Uij=-Uji}) and (\ref{eq:Jac_ident}) lead to
\begin{equation}
(\mathbf{UH})^{t}=\mathbf{H}^{t}\mathbf{U}^{t}=-\mathbf{H}^{t}\mathbf{U}=%
\mathbf{UH},  \label{eq:HU}
\end{equation}
where $\mathbf{U}$ is invertible because the set of operators $S_{N}$ is
linearly independent. Taking into account that $\mathbf{U}^{-1}\mathbf{H}^{t}%
\mathbf{U}=-\mathbf{H}$ we conclude that $P(-\lambda )=\det (\mathbf{H}%
+\lambda \mathbf{I})=0$; that is to say: both $\lambda $ and $-\lambda $ are
eigenvalues.

It follows from (\ref{eq:(H-lambda_I)C=0}) and (\ref{eq:HU}) that
\begin{eqnarray}
\mathbf{H}^{t}\mathbf{UC} &=&-\lambda \mathbf{UC}  \nonumber \\
\mathbf{H}^{\dagger }\mathbf{UC}^{*} &=&-\lambda ^{*}\mathbf{UC}^{*}.
\label{eq:H^t_UC}
\end{eqnarray}

If $H$ has a unitary symmetry given by the unitary operator $S$ such that
\begin{equation}
SO_{i}S^{\dagger }=\sum_{j=1}^{N}s_{ji}O_{j},  \label{eq:SHS^(-1)}
\end{equation}
then it follows from
\begin{equation}
S[H,O_{i}]S^{\dagger }=[H,SO_{i}S^{\dagger }],
\end{equation}
that
\begin{equation}
\mathbf{SH}=\mathbf{HS},  \label{eq:SH_matrix}
\end{equation}
where $\mathbf{S}$ is the matrix with elements $s_{ij}$. Under these
conditions we have
\begin{equation}
\mathbf{HSC}=\mathbf{SHC}=\lambda \mathbf{SC}.  \label{eq:HSC}
\end{equation}
Therefore, if $\lambda _{i}\neq \lambda _{j}$ for all $i\neq j$ then $%
\mathbf{SC}_{i}=s_{i}\mathbf{C}_{i}$.

It follows from
\begin{equation}
S[O_{i},O_{j}]S^{\dagger }=[SO_{i}S^{\dagger },SO_{j}S^{\dagger }]=U_{ij},
\end{equation}
that
\begin{equation}
\mathbf{S}^{t}\mathbf{US}=\mathbf{U}.
\end{equation}

Suppose that $H$ has an antiunitary symmetry given by the antiunitary
operator $A$ and that
\begin{equation}
AO_{i}A^{-1}=\sum_{j=1}^{N}a_{ji}O_{j}.  \label{eq:AOA^(-1)}
\end{equation}
Then, it follows from
\begin{equation}
A[H,O_{i}]A^{-1}=[H,AO_{i}A^{-1}],
\end{equation}
that
\begin{equation}
\mathbf{AH}^{*}=\mathbf{HA},  \label{eq:AH_matrix}
\end{equation}
where $\mathbf{A}$ is the matrix with elements $a_{ij}$. Therefore, if $%
\mathbf{C}$ is an eigenvector of $\mathbf{H}$ with eigenvalue $\lambda $, we
have $\mathbf{HAC}^{*}=\mathbf{AH}^{*}\mathbf{C}^{*}=\mathbf{A}\left(
\mathbf{HC}\right) ^{*}$ that leads to
\begin{equation}
\mathbf{HAC}^{*}=\lambda ^{*}\mathbf{AC}^{*}.  \label{eq:HAC^*}
\end{equation}
We conclude that both $\lambda $ and $\lambda ^{*}$ are eigenvalues of $%
\mathbf{H}$ and roots of the characteristic polynomial $P(\lambda )$. If $%
\mathbf{AC}^{*}=b\mathbf{C}$, where $b$ is a scalar, then the antiunitary
symmetry is exact and $\lambda =\lambda ^{*}$.

Since the set of operators $S_{N}$ is assumed to be linearly independent
then $\mathbf{A}$ is invertible and
\begin{equation}
\mathbf{H}^{*}=\mathbf{A}^{-1}\mathbf{HA}.  \label{eq:A^(-1)HA}
\end{equation}
Besides, it follows from
\begin{equation}
A[O_{i},O_{j}]A^{-1}=[AO_{i}A^{-1},AO_{i}A^{-1}]=U_{ij}^{*},
\end{equation}
that
\begin{equation}
\mathbf{A}^{t}\mathbf{UA}=\mathbf{U}^{*}=\mathbf{U}^{t}=-\mathbf{U}.
\end{equation}

\subsection{Hermitian Hamiltonians}

\label{subsec:Hermitian}

The case $H^{\dagger }=H$ is of particular interest for several physical
applications\cite{SLZEK11,RSEGK12,BGOPY13,POLMGLFNBY14} and was studied in
detail in our earlier papers\cite{F15a,F16}. In what follows we summarize
the main results.

When $H$ is Hermitian we have some additional useful relationships. For
example, the commutator relation $[H,O_{i}]^{\dagger }=-[H,O_{i}]$ leads to
\begin{equation}
H_{ij}^{*}=-H_{ij}.  \label{eq:H*=-H}
\end{equation}
It follows from this equations (\ref{eq:HU}) and (\ref{eq:U^dagger=U}) that $%
\mathbf{H}$ is $\mathbf{U}$-pseudo-Hermitian\cite{F16}:
\begin{equation}
\mathbf{H}^{\dagger }=\mathbf{UHU}^{-1},  \label{eq:H_pseudo_Herm}
\end{equation}
(see \cite{P43,SGH92,M02a,M02b,M02c} for a more general and detailed
discussion of pseudo-Hermiticity or quasi-Hermiticity). Note that $\mathbf{H}
$ and $\mathbf{H}^{\dagger }$ share eigenvalues
\begin{equation}
\mathbf{H}^{\dagger }\mathbf{UC}=\lambda \mathbf{UC.}  \label{eq:H+UC}
\end{equation}

Another important relationship follows from the fact that $[H,Z]^{\dagger
}=-[H,Z^{\dagger }]$:
\begin{equation}
\lbrack H,Z^{\dagger }]=-\lambda ^{*}Z^{\dagger }.  \label{eq:[H,Z+]}
\end{equation}
According to what was argued above if $\left| \psi \right\rangle $ and $%
Z\left| \psi \right\rangle $ belong to the Hilbert space, then both $E$ and $%
\lambda $ are real. In the language of quantum mechanics we often say that $%
Z $ and $Z^{\dagger }$ are a pair of annihilation-creation or ladder
operators because, in addition to (\ref{eq:HZ|Psi>}), we also have
\begin{equation}
HZ^{\dagger }\left| \psi \right\rangle =(E-\lambda )Z^{\dagger }\left| \psi
\right\rangle .  \label{eq:HZ+|Psi>}
\end{equation}

\subsection{Quadratic Hamiltonians}

\label{subsec:QH}

The simplest Hamiltonians that can be treated by the algebraic method are
those that are quadratic functions of the coordinates and their conjugate
momenta:
\begin{equation}
H=\sum_{i=1}^{2K}\sum_{j=1}^{2K}\gamma _{ij}O_{i}O_{j},
\label{eq:H_quadratic}
\end{equation}
where $\left\{ O_{1},O_{2},\ldots ,O_{2K}\right\} =\left\{
x_{1},x_{2},\ldots ,x_{K},p_{1},p_{2},\ldots ,p_{K}\right\} $, $%
[x_{m},p_{n}]=i\delta _{mn}$, and $[x_{m},x_{n}]=[p_{m},p_{n}]=0$. In this
case the matrix $\mathbf{U}$ has the form
\begin{equation}
\mathbf{U}=i\left(
\begin{array}{ll}
\mathbf{0} & \mathbf{I} \\
-\mathbf{I} & \mathbf{0}
\end{array}
\right) ,  \label{eq:U_matrix}
\end{equation}
where $\mathbf{0}$ and $\mathbf{I}$ are the $K\times K$ zero and identity
matrices, respectively, which already satisfies $\mathbf{U}^{\dagger }=%
\mathbf{U}^{-1}=\mathbf{U}$. The matrices $\mathbf{H}$, $\mathbf{\gamma }$
and $\mathbf{U}$ are related by
\begin{equation}
\mathbf{H}=(\mathbf{\gamma }+\mathbf{\gamma }^{t})\mathbf{U,}
\label{eq:H=gamma.U}
\end{equation}
where $\mathbf{\gamma }$ is the matrix with elements $\gamma _{ij}$.

If $\mathbf{\gamma }^{\dagger }=\mathbf{\gamma }$ the quadratic Hamiltonian (%
\ref{eq:H_quadratic}) is Hermitian and $\mathbf{H}$ is $\mathbf{U}$-pseudo
Hermitian: $\mathbf{H}^{\dagger }=\mathbf{UHU}$. In this case $\mathbf{%
\gamma +\gamma }^{t}=\mathbf{\gamma +\gamma }^{*}=2\Re \mathbf{\gamma }$.

The Schr\"{o}dinger equation for a quadratic Hamiltonian is exactly solvable
and its eigenvalues and eigenvectors can be obtained by several approaches%
\cite{RK05,BGOPY13,BG15}. In what follows we apply the algebraic method and
just obtain the eigenvalues $\lambda _{i}$ of the adjoint or regular matrix
representation which is sufficient for present purposes. In this case the
adjoint or regular matrix is closely related to the fundamental matrix\cite
{CGHS12}.

\subsection{One-dimensional example}

\label{subsec:QH1D}

We first consider the simplest quadratic Hamiltonian
\begin{equation}
H=p^{2}+x^{2}+b(xp+px),  \label{eq:H1D}
\end{equation}
which is Hermitian when $b$ is real. On choosing the set of operators $%
S_{2}=\{x,p\}$ we obtain the matrix representation
\begin{equation}
\mathbf{H}=\left(
\begin{array}{ll}
-2ib & 2i \\
-2i & 2ib
\end{array}
\right) .  \label{eq:H1D_mat}
\end{equation}
Its eigenvalues
\begin{equation}
\lambda _{2}=-\lambda _{1}=2\sqrt{1-b^{2}},  \label{eq:lambda_i_1D}
\end{equation}
are real when $b^{2}<1$.

The eigenvalues of the Hermitian case are real when $-1<b<1$ and this
operator does not have eigenvectors in the Hilbert space when $b^{2}\geq 1$.
When $b=1$ the matrix $\mathbf{H}$ exhibits only one eigenvalue $\lambda =0$
and one eigenvector
\begin{equation}
\mathbf{C}=\frac{1}{\sqrt{2}}\left(
\begin{array}{l}
1 \\
1
\end{array}
\right) ,
\end{equation}
and is defective or not diagonalizable. When $b=-1$ we obtain the same
eigenvalue but in this case the only eigenvector is
\begin{equation}
\mathbf{C}=\frac{1}{\sqrt{2}}\left(
\begin{array}{c}
1 \\
-1
\end{array}
\right) ,
\end{equation}
and again the matrix is defective. These particular values of $b$ are
commonly called exceptional points\cite{HS90,H00,HH01,H04}.

In the coordinate representation, the ground-state eigenfunction and
eigenvalue are
\begin{eqnarray}
\psi _{0}(x) &=&Ne^{-\alpha x^{2}},\;\alpha =\frac{1}{2}\left( \sqrt{1-b^{2}}%
+ib\right) ,  \nonumber \\
E &=&\sqrt{1-b^{2}},
\end{eqnarray}
respectively. We clearly appreciate that $\psi _{0}(x)$ is square-integrable
only when $b^{2}<1$ in agreement with the result of the algebraic method.

When $b=i\beta $, the eigenvalues of $\mathbf{H}$ are real for all real
values of $\beta $. In this case there is an antiunitary symmetry $A=PT$
commonly called PT symmetry given by $AH(x,p)A=TH(-x,-p)T=H(-x,p)^{*}$. Its
matrix representation
\begin{equation}
\mathbf{A}=\left(
\begin{array}{cc}
-1 & 0 \\
0 & 1
\end{array}
\right) ,
\end{equation}
satisfies $\mathbf{AA=I}$ and $\mathbf{AHA=H}^{*}$ as argued in section~\ref
{sec:algebraic_method}. Note that since $PH(x,p)P=H(-x,-p)=H(x,p)$ we also
have $AH(x,p)A=H(x,-p)^{*}$. The matrix representation in this case is $-%
\mathbf{A}$ and satisfies exactly the same two conditions.

At any of the exceptional points the adjoint matrix representation can be
written in Jordan canonical form by means of a suitable similarity
transformation. For example, when $b=1$ we have
\begin{equation}
\mathbf{H}=\left(
\begin{array}{ll}
-2i & 2i \\
-2i & 2i
\end{array}
\right) .
\end{equation}
By means of the matrix
\begin{equation}
\mathbf{P}=\left(
\begin{array}{ll}
-2i & 1 \\
-2i & 0
\end{array}
\right) ,
\end{equation}
we obtain
\begin{equation}
\mathbf{P}^{-1}\mathbf{HP}=\left(
\begin{array}{ll}
0 & 1 \\
0 & 0
\end{array}
\right) .
\end{equation}

\section{Two-dimensional examples}

\label{sec:QH2D}

In this section we consider three two-dimensional quadratic Hamiltonians
studied earlier by other authors. The first example
\begin{equation}
H=p_{x}^{2}+p_{y}^{2}+x^{2}+ay^{2}+bxy,\;a>0,  \label{eq:H2D_xy}
\end{equation}
is closely related to the one discussed by Cannata et al\cite{CIN10},
Calliceti et al\cite{CGHS12}, Fern\'{a}ndez and Garcia\cite{FG14b} and Beygi
et al\cite{BKB15} and is Hermitian when $b$ is real. On choosing the basis
set of operators $S_{4}=\{x,y,p_{x},p_{y}\}$ we obtain the regular matrix
representation
\begin{equation}
\mathbf{H}=\left(
\begin{array}{cccc}
0 & 0 & 2i & bi \\
0 & 0 & bi & 2ai \\
-2i & 0 & 0 & 0 \\
0 & -2i & 0 & 0
\end{array}
\right) .
\end{equation}
Its four eigenvalues are the square roots of
\begin{equation}
\xi _{\pm }=2\left[ a+1\pm \sqrt{b^{2}+(a-1)^{2}}\right] ,
\end{equation}
and are real provided that $\xi _{\pm }>0$. More precisely, they are real if
$-(a-1)^{2}<b^{2}<4a$ that reveals four exceptional points, two at each
endpoint. The right exceptional points $b=\pm 2\sqrt{a}$ appear in the
Hermitian case. On the other hand, when $b=i\beta $, $\beta $ real, the
eigenvalues are real if $|\beta |<|a-1|$. In this case the non-Hermitian
operator is similar to a self-adjoint one\cite{CGHS12}. The most symmetric
case $a=1$ exhibits real eigenvalues only in the Hermitian region $0<b^{2}<4$%
.

When $b^{*}=-b$ the operator exhibits two antiunitary symmetries $%
A_{x}=S_{x}T$ and $A_{y}=S_{y}T$, where $%
S_{x}H(x,y,p_{x},p_{y})S_{x}=H(-x,y,-p_{x},p_{y})$ and $%
S_{y}H(x,y,p_{x},p_{y})S_{y}=H(x,-y,p_{x},-p_{y})$, that lead to $%
A_{x}H(x,y,p_{x},p_{y})A_{x}=H(-x,y,p_{x},-p_{y})^{*}$ and $%
A_{y}H(x,y,p_{x},p_{y})A_{y}=H(x,-y,-p_{x},p_{y})^{*}$ with matrix
representations $\mathbf{A}_{x}$ and $\mathbf{A}_{y}$ that satisfy $\mathbf{A%
}_{q}\cdot \mathbf{A}_{q}=\mathbf{I}$ and $\mathbf{A}_{q}\cdot \mathbf{H}%
\cdot \mathbf{A}_{q}=\mathbf{H}^{*}$, $q=x,y$. These antiunitary symmetries
have been named partial parity-time symmetry\cite{BKB15,Y14} and are
particular cases of other antiunitary symmetries that one can find in
quadratic Hamiltonians\cite{F16b}.

The second example was discussed earlier by Li and Miao\cite{LM12} and more
recently by Miao and Xu\cite{MX16} as a three-parameter model but for
present purposes we can rewrite it as a two-parameter one:
\begin{equation}
H=p_{x}^{2}+p_{y}^{2}+x^{2}+ay^{2}+bp_{x}p_{y},\;a>0.  \label{eq:H2D_pxpy}
\end{equation}
In those articles $b$ was chosen to be imaginary but here we allow it to be
also real, in which case the Hamiltonian is Hermitian.

The four eigenvalues of its regular matrix representation
\begin{equation}
\mathbf{H}=\left(
\begin{array}{cccc}
0 & 0 & 2i & 0 \\
0 & 0 & 0 & 2ai \\
-2i & -bi & 0 & 0 \\
-bi & -2i & 0 & 0
\end{array}
\right) ,
\end{equation}
are the square roots of
\begin{equation}
\xi _{\pm }=2\left[ a+1\pm \sqrt{ab^{2}+(a-1)^{2}}\right] .
\end{equation}
These eigenvalues are real provided that
\begin{equation}
-\frac{\left( a-1\right) ^{2}}{a}<b^{2}<4,
\end{equation}
which reveals that there are four exceptional points as in the preceding
example. The right ones $b=\pm 2$ appear in the Hermitian case. On the other
hand, when $b=i\beta $, $\beta $ real, the eigenvalues are real if $|\beta
|<|a-1|/\sqrt{a}$. The most symmetric case $a=1$ exhibits real eigenvalues
only in the Hermitian region $0<b^{2}<4$.

Clearly, most of the mathematical features of the models (\ref{eq:H2D_xy})
and (\ref{eq:H2D_pxpy}) are similar. To what has just been said we add that
when $b^{*}=-b$ both operators exhibit exactly the same two antiunitary
symmetries $A_{x}$ and $A_{y}$ already discussed above.

According to Li and Miao\cite{LM12} and Miao and Xu\cite{MX16} when $%
b^{*}=-b $ the operator (\ref{eq:H2D_pxpy}) is neither Hermitian nor PT
symmetric but it is PT-pseudo Hermitian because $PTHPT=H^{*}=H^{\dagger }$.
However, an operator $H$ is $\eta $-pseudo Hermitian if there is an
invertible Hermitian operator $\eta $ such that $H^{\dagger }=\eta H\eta
^{-1}$\cite{P43,SGH92,M02a,M02b,M02c}. Since the operator $PT$ is not
Hermitian (it is in fact antilinear and antiunitary) it is not correct to
speak of PT-pseudo Hermiticity. The operator (\ref{eq:H2D_pxpy}) exhibits
two antiunitary symmetries given by $A_{x}$ and $A_{y}$ as well as the true
pseudo-Hermiticity provided by an operator $\tilde{\eta}=\tilde{\Omega}%
^{\dagger }\tilde{\Omega}$, where $\tilde{\Omega}$ is a suitable exponential
operator\cite{MX16}.

The next example is a pair of harmonic oscillators coupled by an angular
momentum and was proposed for the study of ``a particle in a rotating
anisotropic harmonic trap or a charged particle in a fixed harmonic
potential in a magnetic field''\cite{RCR14}. For simplicity we rewrite this
Hamiltonian in the following form
\begin{equation}
H=p_{x}^{2}+p_{y}^{2}+x^{2}+ay^{2}+b\left( xp_{y}-yp_{x}\right) ,\;a>0.
\label{eq:H2D_xpy-ypx}
\end{equation}
It is Hermitian when $b$ is real which is exactly the case considered by
Reb\'{o}n et al\cite{RCR14} but here we also allow it to be complex.

The regular matrix representation for this operator is
\begin{equation}
\mathbf{H}=\left(
\begin{array}{cccc}
0 & -bi & 2i & 0 \\
bi & 0 & 0 & 2ai \\
-2i & 0 & 0 & -bi \\
0 & -2i & bi & 0
\end{array}
\right) ,
\end{equation}
and its eigenvalues are the square roots of
\begin{equation}
\xi _{\pm }=2a+b^{2}+2\pm \sqrt{(a-1)^{2}+2(a+1)b^{2}}.
\end{equation}
These eigenvalues are real provided that
\begin{eqnarray}
b^{2} &>&-\frac{(a-1)^{2}}{2(a+1)},  \nonumber \\
0 &>&\left( b^{2}-4\right) \left( 4a-b^{2}\right) .
\end{eqnarray}
Once again we appreciate that when $a=1$ the eigenvalues are real only when
the Hamiltonian is Hermitian ($-2<b<2$).

The Hamiltonian (\ref{eq:H2D_xpy-ypx}) is parity invariant and when $%
b^{*}=-b $ it is also PT symmetric: $%
PTHPT=TH(x,y,p_{x},p_{y})T=H(x,y,-p_{x},-p_{y})^{*}=H(x,y,p_{x},p_{y})$. The
matrix representation $\mathbf{A}$ of this antiunitary symmetry satisfies $%
\mathbf{A}\cdot \mathbf{A}=\mathbf{I}$ and $\mathbf{A}\cdot \mathbf{H}\cdot
\mathbf{A}=\mathbf{H}^{*}$ as argued above.

The characteristic polynomial for the three examples discussed in this
section is
\begin{equation}
\lambda ^{4}-\left( \xi _{+}+\xi _{-}\right) \lambda ^{2}+\xi _{+}\xi _{-}=0,
\end{equation}
therefore, the dynamical variables satisfy the differential equation of
fourth order
\begin{equation}
\frac{d^{4}}{dt^{4}}q+\left( \xi _{+}+\xi _{-}\right) \frac{d^{2}}{dt^{2}}%
q+\xi _{+}\xi _{-}=0,
\end{equation}
as shown in the Appendix.

The equations above show that $\xi _{\pm }=\omega ^{2}\pm \Delta $, so that
the case of equal frequencies takes place at the exceptional points given by
the condition $\Delta =0$. In this case the corresponding adjoint matrix
representation $\mathbf{H}_{\omega }$ can be written in Jordan canonical
form
\begin{equation}
\mathbf{PH}_{\omega }\mathbf{P}=\left(
\begin{array}{cccc}
-\omega & 1 & 0 & 0 \\
0 & -\omega & 0 & 0 \\
0 & 0 & \omega & 1 \\
0 & 0 & 0 & \omega
\end{array}
\right) ,
\end{equation}
where the form of the matrix $\mathbf{P}$ depends on the model.

\section{Further comments and conclusions}

\label{sec:conclusions}

Quadratic Hamiltonians are suitable models for the study of several physical
phenomena\cite{SLZEK11,BGOPY13,POLMGLFNBY14} as well as theoretical
investigations\cite{E80,BG15,RCR14}. The algebraic method is a suitable tool
for the analysis of the spectra of such oscillators\cite{F15a,F16}. In this
paper we extended the treatment from Hermitian quadratic Hamiltonians to
non-Hermitian ones taking into account possible unitary and antiunitary
symmetries.

A common feature of the two-dimensional oscillators discussed in section~\ref
{sec:QH2D} is that they do not exhibit real eigenvalues when $a=1$ and $%
b^{2}<0$. According to the algebraic method an eigenvalue $\lambda $ of the
regular matrix representation $\mathbf{H}$ is the difference between two
energy levels of the Hamiltonian $H$. Therefore this frequency reveals the
dependence of the energy levels on the model parameter $b$. For this reason,
when $a\neq 1$ the energy levels can be expanded in a Taylor series of the
form
\begin{equation}
E_{n_{1}n_{2}}=\sum_{j=0}^{\infty
}E_{n_{1}n_{2}}^{(j)}b^{j},\;E_{n_{1}n_{2}}^{(2j+1)}=0,
\end{equation}
and the eigenvalues are real for $b^{2}<0$ within its radius of convergence.
If, on the other hand, $a=1$ the perturbation corrections of odd order do
not vanish and the eigenvalues can only be real for real $b$. In a series of
papers we proposed to calculate the perturbation correction of first order
to determine whether the eigenvalues of a given non-Hermitian operator are
real or complex\cite{FG14a,FG14b,AFG14b,AFG14c}. If the perturbation
correction of first order is nonzero, then the eigenvalues are complex. This
simple and straightforward argument applies even if the perturbation series
is divergent as long as it is asymptotic to the actual eigenvalue. Whether
the perturbation correction of first order vanishes or not depends on the
symmetry of $H_{0}$; for this reason the application of group theory proved
to be quite successful\cite{FG14a,FG14b,AFG14b,AFG14c}. The quadratic
Hamiltonians studied in section~\ref{sec:QH2D} are exactly solvable problems
that confirm the argument put forward in those earlier papers. Note that the
greater symmetry of $H_{0}$ takes place when $a=1$.

\section*{Appendix}

In this appendix we derive some additional results that may be useful for
future applications of the algebraic method.

For every eigenvalue $\lambda _{i}$ we construct the operator
\begin{equation}
Z_{i}=\sum_{j=1}^{2K}c_{ij}O_{j}.
\end{equation}
For convenience we label the eigenvalues in such a way that $\lambda
_{j}=-\lambda _{2K-j+1}$, $j=1,2,\ldots ,K$, and when they are real we
organize them in the following way:
\begin{equation}
\lambda _{1}<\lambda _{2}<\ldots <\lambda _{K}<0<\lambda _{K+1}<\ldots
<\lambda _{2K}.
\end{equation}

If we take into account that $[H,Z_{i}Z_{j}]=\left( \lambda _{i}+\lambda
_{j}\right) Z_{i}Z_{j}$ then we conclude that
\begin{equation}
\lbrack H,[Z_{i},Z_{j}]]=\left( \lambda _{i}+\lambda _{j}\right)
[Z_{i},Z_{j}]=0,
\end{equation}
which tells us that $Z_{i}$ and $Z_{j}$ commute when $\lambda _{i}+\lambda
_{j}\neq 0$. If $[Z_{j},Z_{2K-j+1}]=\sigma _{j}\neq 0$ for all $j=1,2,\ldots
,K$ then we can write $H$ in the following way
\begin{equation}
H=-\sum_{j=1}^{K}\frac{\lambda _{j}}{\sigma _{j}}Z_{2K-j+1}Z_{j}+E_{0}.
\end{equation}

If $\psi _{0}$ is a vector in the Hilbert space where $H$ is defined that
satisfies
\begin{equation}
Z_{j}\psi _{0}=0,\;j=1,2,\ldots ,K,
\end{equation}
then $H\psi _{0}=E_{0}\psi _{0}.$

Consider the time-evolution of the dynamical variables
\begin{equation}
O_{j}(t)=e^{itH}O_{j}e^{-itH},
\end{equation}
so that
\begin{equation}
\dot{O}_{j}(t)=ie^{itH}[H,O_{j}]e^{-itH}=i\sum_{k=1}^{2K}H_{kj}O_{k}(t).
\end{equation}
If we define the row vector $\mathbf{O}(t)=\left( O_{1}(t)\,O_{2}(t)\,\ldots
\,O_{2K}(t)\right) $ then we have the matrix differential equation $\mathbf{%
\dot{O}}(t)=i\mathbf{O}(t)\mathbf{H}$ with the following solution:
\begin{equation}
\mathbf{O}(t)=\mathbf{O}e^{it\mathbf{H}},\;\mathbf{O}=\mathbf{O}(0).
\end{equation}
Since $P(\mathbf{H})=0$ then
\begin{equation}
P\left( -i\frac{d}{dt}\right) \mathbf{O}(t)=\mathbf{O}P(\mathbf{H})e^{it%
\mathbf{H}}=0,
\end{equation}
gives us a differential equation of order $2K$ for the dynamical variables.
Obviously, $Z_{j}(t)=e^{it\lambda _{j}}Z_{j}$, $j=1,2,\ldots ,2K$, satisfies
this equation.

\end{document}